\documentclass[11pt]{article}

\usepackage{amssymb,amsmath,latexsym}
\usepackage{undertilde}
\usepackage[totalwidth=17cm,totalheight=24cm]{geometry}
\newcommand{\C}{{\mathbb C}}

\newcommand{\im}{{\rm i }}

\newcommand\be{\begin{eqnarray}}
\newcommand\ee{\end{eqnarray}}

\begin{document}

\title{Fermions via spinor-valued one-forms}
\author{Alexander Torres-Gomez$^{a,b}$ and Kirill Krasnov$^b$\\
\textit{\small ${}^a$ Instituto de Ciencias Fisicas y Matematicas, Universidad Austral de Chile, Valdivia, Chile} \\
\textit{\small ${}^b$ School of Mathematical Sciences, University of Nottingham, Nottingham, UK}}
\date{December 2012}
\maketitle

\begin{abstract} Spinor-valued one-forms (Rarita-Schwinger fields) are normally used in the context of supergravity, where they describe spin $3/2$ particles (gravitinos). Indeed, when decomposed into irreducible representations of the Lorentz group such a field contains both a spin $1/2$ and a spin $3/2$ component, and the Rarita-Schwinger Lagrangian is designed to make only the spin $3/2$ propagate. We point out that the opposite construction is also possible, and give a spinor-valued one-form field Lagrangian that describes a propagating spin $1/2$ particle. 
\end{abstract}

\section{Introduction}

Our current description of Nature operates with three different types of fields. First, there is the gravitational field, to which all other fields couple universally. Second, there are bosonic gauge fields and the Higgs. Third, there are fermions. The Lagrangians used to encode the dynamics of these fields are all quite different. Thus,  for gravity we use the non-polynomial, but well-motivated from the geometric point of view Einstein-Hilbert Lagrangian. It leads to second order in derivatives field equations. Then, the bosonic ingredients of the Standard Model are described by a polynomial (renormalizable) Lagrangian, with again second order field equations. Finally, fermions are described by the Dirac Lagrangian, with first order field equations. 

We are used to the fact that gravity is so different from the rest of the interactions because it is universal, and as such can be encoded by the very geometry of space and time. We also got used to the fact that it is the only non-renormalizable interaction. On the other hand, it may appear that fermions are different just because they are described by first, instead of second order in derivatives Lagrangian. This is of course incorrect, as can be seen from the fact that all other interactions, with introduction of additional fields, can be rewritten as first order systems. Thus, in the case of gravity this can be achieved by introducing the connection as an independent variable (incidentally, this also makes the Lagrangian polynomial, which shows that non-polynomiality by itself is not the cause of problems with gravity). In the case of Yang-Mills fields the first order formulation can be obtained by re-writing the Yang-Mills Lagrangian in the so-called BF form. The opposite is also possible, and some of the fields of the first order Dirac Lagrangian can be integrated out to produce a second order formulation of fermions, see more on this below. Thus, the order of the field equations is, at least to some extent, just a matter of convenience of the description. Both second and first order formulations are generally possible for any given system, and the difference between the two is often just the difference between the Lagrangian and Hamiltonian formulations. 

The real difference between the fermions and all other fields lies in their spin and statistics. The latter makes it most natural to use anti-commuting Grassmann variables to describe fermions. If this is the principal difference, one can ask if fermions can be described by Lagrangians of the same type as those used for bosons, just with the Grassmann-valued fields used. To a certain extent this is possible, and the subject of the present paper is to study some of these issues. 

As we have already mentioned, and as will be reviewed below, fermions can be described by second order in derivatives Lagrangians. This brings them closer to the standard description of the bosonic fields. However, for reasons to be explained below, we would like to do more and describe the usual spin 1/2 Dirac fermions using spinor and Grassmann-valued {\it one-forms} as the basic fields. Rephrasing, we try to describe the Dirac fermions as a sort of gauge-fields, but corresponding to anti-commuting gauge group generators. As we shall try to convince the reader, such a description is at least to some extent possible (we will describe its difficulties after we present the construction). 

A motivation for our construction comes from the fact that it is possible to describe gravity using a gauge field instead of the metric as the basic field \cite{Krasnov:2011pp}. Oversimplifying, the idea is as follows. We have already mentioned that there exists a first order formulation of gravity with the connection as an independent variable. As is often the trick with the first order formulations, one can integrate out the original field (the metric) and obtain a new second-order Lagrangian that is a functional of only the connection. If one does this to the Palatini first order formulation, one obtains the theory proposed and studied long time ago by Eddington \cite{Eddington}. If one performs the same with the so-called Plebanski formulation \cite{Plebanski:1977zz} of GR, one obtains the formulation \cite{Krasnov:2011pp}. In the latter version, the resulting gauge-theoretic formulation of gravity exhibits many similarities with Yang-Mills theory, see \cite{Krasnov:2012pd} for more details. Moreover, in this framework both the gravitational and gauge boson degrees of freedom can be put together in a larger connection field, with part describing gravity and another part describing Yang-Mills fields, see \cite{Krasnov:2012pd,Krasnov:2011hi} for more details. In this approach it appears to be most natural to attempt to add fermions just by making the connection field even larger, so that its components corresponding to anti-commuting generators describe particles with half-integer spin. There is no guarantee that this is possible, and this paper is a preliminary step in this direction. 

It may be objectionable to many readers to describe fermions as components of a connection, even if Grassmann-valued. Indeed, we are used to the particle physics picture of fermions being described by the fundamental representations of the corresponding gauge groups, while gauge bosons are charged under the adjoint. How can both of these be put together into a single object? However, this objection can be overcome in the framework of Lie superalgebras. Indeed, the basic definition of a Lie superalgebra is that of a graded vector space with a (super)-commutator, such that the subspace of odd elements forms a representation of the even sub-algebra, see e.g. \cite{Frappat:1996pb} for a useful description. Thus, there is no formal problem in putting together objects that transform under some representation of the gauge group (fermions) with the objects that act on them (gauge bosons), with Lie superalgebras achieving exactly this. However, the fact that this is in principle possible does not guarantee that it is possible to do this in a physically realistic fashion. This paper is a step towards understanding how far one can get with this idea. Related ideas in the context of 2+1 gravity were explored in \cite{Alvarez:2011gd}.

With these motivating remarks being made, the Lagrangian that we propose is as follows. Let us for simplicity concentrate on the case of a single Majorana fermion (electrically uncharged). The Dirac case is treated in the main text. We use a single spinor- and Grassmann-valued one-form field $\rho_\mu^A$, where $A, B, \ldots=1,2$ is our notation for the 2-component spinor index, and $\mu$ is the spacetime index. We assume the Minkowski spacetime background with metric $\eta_{\mu\nu}$. The action will also explicitly contain the self-dual two-forms $\Sigma_{\mu\nu}^{AB}$, where $AB=(AB)$ is a symmetric pair of spinor indices. An explicit expression for $\Sigma_{\mu\nu}^{AB}$ in terms of the soldering form is given below, see (\ref{Sigma-def}). The Lagrangian reads:
\begin{equation}\label{intr-L}
{\cal L}= - 2  (\Sigma^{\mu\nu\,AC}\partial_\mu \rho_{\nu\,C})^2- \frac{3m^2}{2}  (\rho_{\mu}^A)^2.
\end{equation}
The numerical factor in front of the first term is introduced for future convenience. The spinor indices here are contracted with the help of the spinor metric $\epsilon_{AB}$, and the spacetime index in the last term using the metric $\eta_{\mu\nu}$. We take our spinor-valued one-form $\rho_\mu^A$ to have the mass dimension one $[\rho]=1$, and thus the first term in the Lagrangian has the required mass dimension 4. The last term then contains a dimensionful constant $m^2$ of dimensions mass squared. The main claim of the paper is that the above Lagrangian describes a single (uncharged) spin $1/2$ particle of mass $m$. 

The above Lagrangian clearly leads to second order field equations. Also, unlike the case with Majorana/Weyl or Dirac Lagrangians, no hermitian conjugate fields appear in (\ref{intr-L}), and so it is not Hermitian. The immediate question is then how can such a Lagrangian be equivalent to the first order Hermitian Majorana Lagrangian. To explain this, we need to start with some remarks. 

First, we note that many textbooks describe fermions using only the technology of 4-component fermions and $4\times 4$ $\gamma$-matrices. However, it has been appreciated for quite some time that working with 2-component fermions is conceptually more clear (even though not very practical for things like e.g. QED Feynman diagram computations). A rather complete description of fermions via 2-component spinors is given in e.g. \cite{Dreiner:2008tw}. This reference also describes the Standard Model in the 2-component fermion language. For a textbook treatment that uses 2- as well as 4-component spinors see e.g. \cite{Srednicki:2007qs}. 

When fermionic Lagrangians are written in the 2-component form, an interesting possibility arises. This Lagrangian, being Hermitian, necessarily involves 2-component spinors in both fundamental representations of the Loretz group, i.e. unprimed $(1/2,0)$ and primed $(0,1/2)$ ones (we are using the GR community terminology here, instead of undotted and dotted spinors common in the particle physics literature). At the "classical" level of the field equations the primed spinors are required to be the complex (or Hermitian) conjugates of the corresponding unprimed spinors, and so they are not independent objects (and this "reality condition" ensures hermiticity of the Lagrangian and thus unitarity). However, at the level of the path integral the fermionic fields of opposite chiralities are integrated over independently. One could try to mimic what happens in the path integral already at the level of the Lagrangian, and integrate out all spinor fields of one type to obtain a purely chiral Lagrangian. It is not hard to see that it will be second-order in derivatives. 

Such a second-order chiral formulation  of fermions has been proposed in particular in \cite{Chalmers:1997ui}, see also \cite{Morgan:1995te} for an earlier reference. The work  \cite{Chalmers:1997ui} also emphasized the important simplifications that occur in this formalism as compared to the usual first-order one. As the authors point out, much of the algebra of $\gamma$-matrices needed when computing with the usual formalism has been done once and for all by the procedure of integrating out the primed spinors. This results in simplifications in both propagators and interaction vertices. In this second-order formulation the 2-component description of fermions actually becomes more efficient for computing Feynman diagrams than the original Dirac description. Admittedly, some aspects (such as e.g. unitarity) become less manifest in the chiral description, but the simplicity of the formalism is worth the price. Some aspects of this not widely known formalism will be reviewed below. 

To summarize, it is possible to rewrite the usual first order in derivatives Hermitian Lagrangian for fermions in a second order form, which also makes the Hermiticity not manifest. Our Lagrangian (\ref{intr-L}) is similar, and in this respect is not new. What is new is that, for reasons already explained above, we decided to describe our fermion using a spinor-valued one-form instead of a spinor-valued function. Such objects are familiar from the supergravity literature, where they go under the name of Rarita-Schwinger fields, and are used to described spin $3/2$ particles. In contrast, our Lagrangian is designed in such a way that only the $1/2$ component of the spinor-valued one-form propagates. Our main objective in this paper is to verify the propagating mode content of (\ref{intr-L}), and present some generalizations.

The organisation of the paper is as follows. Some basic facts about 2-component spinors, in the amount we need, are reviewed in the Appendix. In Section \ref{sec:fermions} we start by giving a description of the usual Weyl and Dirac fermions in the language of 2-component spinors.  In this section we also remind the reader how fermions can be described using a second order in derivatives formulation. The corresponding Hamiltonian formulations are reviewed in the Appendix. Then, in Section \ref{sec:RS} we  review the Rarita-Schwinger Lagrangian, in the language of 2-component spinors.  Section \ref{sec:maj} then gives a description of a single massive Majorana fermion via a spinor-valued one-form field. Section \ref{sec:dirac} generalizes this to the case of a Dirac fermion. 

Let us note that unless otherwise specified, all rank one spinors that we consider in this paper are Grassmann valued, i.e. their components are anti-commuting. Our signature is $(-,+,+,+)$.

\section{Preliminaries: Second order formulation of fermions}
\label{sec:fermions}

\subsection{A single massless Weyl fermion}

The Lagrangian for a single massless Weyl fermion reads:
\be\label{weyl-massless}
{\cal L}_{\text{Weyl}}= -\im \sqrt{2} (\lambda^\dagger)_{A'} \theta^{\mu\,A'A} \partial_\mu \lambda_A\equiv -\im \sqrt{2} \lambda^\dagger \theta^{\mu}\partial_\mu \lambda.
\ee
Here $\lambda_A$ is a 2-component spinor, $\lambda_A^{\dagger}$ is its Hermitian conjugate  and $\theta_\mu^{A'A}$ is the soldering form, see (\ref{theta-io}) for an explicit expression. We have also written the Lagrangian in an index-free way. The factor of $\sqrt{2}$ is introduced for future convenience, and the minus in front of the kinetic term is convention dependent. With our conventions it is needed to get the positive-definite Hamiltonian. It is assumed that the background spacetime is the Minkowski one, and so the usual derivative can be used. Using the Hermitian property of the soldering form, as well as the Grassmann nature of the fermions, one easily checks that the above Lagrangian is Hermitian (modulo a surface term). 

\subsection{The Majorana mass term}

Let us now consider the massive case. Since our fermions are Grassmann valued we can have the Majorana mass term. Thus, consider
\be\label{majorana}
{\cal L}_{\text{Majorana}}=-\im \sqrt{2} \lambda^\dagger \theta^{\mu} \partial_\mu \lambda - (m/2)\lambda\lambda - (m/2)\lambda^\dagger\lambda^\dagger,
\ee
where we have used the index-free notation, and $m$ is the parameter with dimensions of mass, later to be identified with the physical mass. Note that we need to add both terms in order for the Lagrangian to be Hermitian.

\subsection{A chiral formulation for a Majorana fermion}

As we have already mentioned in the introduction, a chiral formulation can be obtained by integrating out all primed fields. In this case this is the $(\lambda^\dagger)^{A'}$ fermionic field, in which the action is quadratic. At first sight it might seem that it is not legitimate to do this, as the field $(\lambda^\dagger)^{A'}$ is not independent from $\lambda_A$, being the conjugate of the latter. However, in the Berezin integration over Grassmann spinors $(\lambda^\dagger)^{A'}$ and $\lambda_A$ {\it are} treated as independent. Thus, it is a legitimate operation to integrate out $(\lambda^\dagger)^{A'}$ at the level of the path integral. The arising action for $\lambda_A$ will not be Hermitian, however, unless some reality conditions are imposed. 

Let us carry out this simple exercise. The field equation that one gets for $(\lambda^\dagger)^{A'}$ is
\be
\im \sqrt{2} \theta^{\mu\,AA'} \partial_\mu \lambda_A  + m(\lambda^\dagger)^{A'}=0,
\ee
from which we find:
\be\label{eqn-dirac}
(\lambda^\dagger)^{A'} = - \frac{\im\sqrt{2}}{m} \theta^{\mu\, A'A} \partial_\mu\lambda_A.
\ee
We now substitute this back into (\ref{majorana}) and get a chiral action involving only $\lambda_A$. We have
\be
{\cal L}_{\text{chiral}}=-\frac{1}{m}\theta_{A'}^{\mu\, A} \partial_\mu \lambda_A \theta^{\nu\,A'B} \partial_\nu \lambda_B - \frac{m}{2}\lambda^A\lambda_A.
\ee
Let us now use the first identity in (\ref{S-theta-id}). The second term produced is anti-symmetric in $\mu\nu$, and so using the possibility to integrate by parts and the fact that partial derivatives commute we see that there is only a contribution from the first term. When the derivative acting on the fermion field is promoted into a covariant derivative the $\Sigma$-term will give rise to an additional term containing the curvature. But in our free fermion case we get the following chiral Lagrangian
\be\label{weyl-chiral}
{\cal L}_{\text{chiral}}=-\frac{1}{2m}\partial^\mu \lambda^A \partial_\mu \lambda_A - \frac{m}{2}\lambda^A\lambda_A,
\ee
which is just the obvious second-order Lagrangian leading to 
\be
(\partial^\mu\partial_\mu - m^2)\lambda^A=0
\ee
 as its field equation.

As it stands, the Lagrangian (\ref{weyl-chiral}) is not Hermitian, and so this theory is not a good starting point for quantization. However, it can be supplemented with a reality condition that makes it completely equivalent to the original first-order theory. Thus, we can treat the "Dirac" equation (\ref{eqn-dirac}) as a reality condition. This selects a real slice of the phase space of the theory (\ref{weyl-chiral}), and on this real section one gets dynamics with a Hermitian Hamiltonian. All in all, the second-order formulation (\ref{weyl-chiral}), supplemented with the reality condition (\ref{eqn-dirac}) is an equally legitimate viewpoint on the Majorana fermion. The simplifications then come from the fact that in computing the Feynman amplitudes one only has to worry about the reality condition on the external lines of the diagrams. For the internal lines the path integral treats $\lambda^A$ and $\lambda^{\dagger A'}$ as independent. Then some of the algebra of $\gamma$-matrices needed for computing Feynman amplitudes has already been done at the level of the action, which results in significant simplifications for practical computations, see \cite{Chalmers:1997ui}. 

\subsection{Dirac fermions}

Dirac fermions are obtained by taking two massive Weyl fermions of equal mass. The system is then invariant under ${\rm SO}(2)$ rotations mixing the fermions. Since ${\rm SO}(2)\sim {\rm U}(1)$, complex linear combinations of fermions can be introduced and the Lagrangian rewritten in an explicitly ${\rm U}(1)$-invariant way:
\be\label{dirac}
{\cal L}_{\text{Dirac}}= -\im \sqrt{2} \xi^\dagger \theta^{\mu} \partial_\mu \xi-\im \sqrt{2} \chi^\dagger \theta^{\mu}\partial_\mu \chi - m\chi\xi -m\xi^\dagger \chi^\dagger.
\ee
We note that unlike the Majorana mass (\ref{majorana}), the Dirac mass terms (the last two terms in the Lagrangian) can be written for both commuting as well as Grassmann fermion fields.

It is obvious that the Lagrangian has the following global ${\rm U}(1)$ symmetry:
\be
\xi \to e^{\im \varphi} \xi, \qquad \chi \to e^{-\im\varphi} \chi.
\ee
This symmetry can be made local by introducing a ${\rm U}(1)$ gauge field and converting the usual derivative to the covariant one. Thus, we replace
\be\label{cov-ders}
\partial_\mu \xi \to D_\mu \xi = (\partial_\mu - \im A_\mu)\xi, \qquad
\partial_\mu \chi \to D_\mu \chi = (\partial_\mu + \im A_\mu)\chi,
\ee
where $A_\mu$ is the electromagnetic potential. Note that, since the fields $\xi,\chi$ are charged in the opposite way, the expressions for the covariant derivatives on these fields differ by a sign in front of $A_\mu$. The gauge transformation rule for the electromagnetic potential is $A_\mu\to A_\mu+\partial_\mu \varphi$. The Lagrangian becomes
\be\label{dirac-a}
{\cal L}_{\text{Dirac}}=- \im \sqrt{2} \xi^\dagger \theta^{\mu} D_\mu \xi-\im \sqrt{2} \chi^\dagger \theta^{\mu}D_\mu \chi - m\chi\xi -m\xi^\dagger \chi^\dagger.
\ee
This is the way that Dirac fermions couple to the electromagnetic potential.

\subsection{A chiral Dirac theory}

As for Weyl fermions considered above, at the level of the path integral we can integrate out the fermionic fields $\xi^\dagger, \chi^\dagger$ and obtain a chiral Lagrangian involving only unprimed spinors. From field equations for the primed spinors we get:
\be\label{eqs-dirac}
(\xi^\dagger)^{A'} = -\frac{\im\sqrt{2}}{m} \theta^{\mu\, A'A} D_\mu\chi_A, \qquad (\chi^\dagger)^{A'} = -\frac{\im\sqrt{2}}{m} \theta^{\mu\, A'A} D_\mu\xi_A.
\ee
Substituting this into the Lagrangian (\ref{dirac-a}) we get:
\be
{\cal L}_{\text{chiral}} = - \frac{2}{m}\theta_{A'}^{\mu\, A} D_\mu \chi_A \theta^{\nu\,A'B} D_\nu \xi_B - m \chi^A\xi_A.
\ee
We now again use the first identity in (\ref{S-theta-id}) to rewrite this Lagrangian as:
\be
{\cal L}_{\text{chiral}} =- \frac{1}{m} D^\mu \chi^A D_\mu \xi_A - m \chi^A\xi_A - \frac{\im}{m} \Sigma^{\mu\nu\,AB}\chi_A  \xi_B F_{\mu\nu},
\ee
where we have integrated by parts to get the last term and $F_{\mu\nu}=2\partial_{[\mu} A_{\nu]}$. The last term describes interactions with the gauge field and can be seen to be essentially the spin to electromagnetic potential coupling term of Pauli's phenomenological description of spin. Note, however, that there are also interaction with the electromagnetic field vertices hidden in the first term. We can further simplify this Lagrangian by rescaling the fields. It is clear that in this formalism it is natural to introduce fermionic fields of mass dimension one via $\chi\to\sqrt{m}\chi, \xi\to\sqrt{m}\xi$. In terms of the rescaled fields the Lagrangian takes a particularly simple form:
\be\label{Dirac-2-order}
{\cal L}_{\text{chiral}} =- D^\mu \chi^A D_\mu \xi_A - m^2 \chi^A\xi_A - \im \Sigma^{\mu\nu\,AB} \chi_A  \xi_B F_{\mu\nu}.
\ee
When supplemented with the "reality conditions" (\ref{eqs-dirac}), this second-order Lagrangian gives an equivalent, but more economic description of the Dirac fermions.

\section{Preliminaries: Rarita-Schwinger field}
\label{sec:RS}

For completeness, before considering a spinor-valued one-form description of a spin 1/2 field, we start with a more standard material on the spin 3/2. A treatment in terms of 4-component spinors can be found in e.g. \cite{Weinberg:2000cr}, see page 335. We give a description in terms of 2-component spinors.  

\subsection{First order description}

For simplicity, we consider a single uncharged spin 3/2 field. It can be described by a single spinor- and Grassmann-valued one-form $\chi^A_\mu$ and its conjugate $\chi^{\dagger A'}_\mu$. The Lagrangian is
\be\label{RS}
{\cal L}_{3/2} = \sqrt{2} \epsilon^{\mu\nu\rho\sigma} \chi_{\mu A'}^\dagger \theta_\nu^{AA'} \partial_\rho \chi_{\sigma A} - m \Sigma^{\mu\nu AB} \chi_{\mu A}\chi_{\nu B} - m \bar{\Sigma}^{\mu\nu A'B'} \chi^\dagger_{\mu A'}\chi^\dagger_{\nu B'}.
\ee
Here $\bar{\Sigma}_{\mu\nu}^{A'B'} = - (\Sigma_{\mu\nu}^*)^{A'B'}$ is the anti-selfdual two-form conjugate of $\Sigma$, and the Lagrangian is Hermitian, modulo a surface term. To see that it describes a spin $3/2$ field, let us write down the field equations. We get
\be\label{RS-feqs}
\sqrt{2}\epsilon^{\mu\nu\rho\sigma} \theta_\nu^{AA'} \partial_\rho \chi_{\sigma A} = 2m\bar{\Sigma}^{\mu
\nu A'B'} \chi_{\nu B'}^\dagger, \qquad 
\sqrt{2} \epsilon^{\mu\nu\rho\sigma} \theta_\nu^{AA'} \partial_\rho \chi^\dagger_{\sigma A'} = -2m\Sigma^{\mu
\nu AB} \chi_{\nu B},
\ee
where the second equation is the complex conjugate of the first one. Taking the divergence of the second equation we see that
\be\label{RS-1}
\Sigma^{\mu\nu AB} \partial_\mu \chi_{\nu B}=0.
\ee
Multiplying the first equation by $\theta_{\mu A'}{}^E$, and using the identities
\be\label{S-theta-id}
\theta_\mu^{AA'} \theta_{\nu A'}{}^B = \frac{1}{2} \epsilon^{AB} \eta_{\mu\nu} - \Sigma^{AB}_{\mu\nu}, \qquad
\bar{\Sigma}^{\mu\nu A'B'} \theta_{\mu A'}{}^E = -\frac{3}{2} \theta^{\nu EB'}
\ee
that follow from the definition (\ref{Sigma-def}), we get, using the fact that $\Sigma_{\mu\nu}^{AB}$ are self-dual
\be
2\im \Sigma^{\mu\nu AE} \partial_\mu \chi_{\nu A} = -3m \theta^{\nu B'E} \chi^\dagger_{\nu B'}.
\ee
But then using (\ref{RS-1}) we see that 
\be\label{RS-2}
 \theta^{\mu A'E} \chi^\dagger_{\mu A'}=0.
 \ee
To see what this implies, let us define certain projector operators. 
 
 \subsection{Projectors}

Using the soldering form $\theta_\mu^{AA'}$ one can convert the spacetime index of $\chi_\mu^A$ into a pair of spinor indices of opposite types. Thus, we get an object $\chi_{MM'}^A$. This object transforms as $S_+\otimes S_+\otimes S_-$ representation of the Lorentz group, where $S_+$ stands for unprimed spinors and $S_-$ for the primed ones. Thus, this object is not irreducible with respect to the action of the Lorentz group. Its two irreducible components are 
\be
\chi^{AMM'} \to \chi^{(AM)M'} \in S_+^2\otimes S_-, \qquad \chi^{M'}:= \chi^{E}{}_{E}{}^{M'} \in S_-.
\ee
The above decomposition can be made explicit with the use of projectors
\be
P_{1/2}^{\mu\nu AB} = \frac{1}{4} \left( \eta^{\mu\nu} \epsilon^{AB} - 2\Sigma^{\mu\nu AB}\right), \qquad
P_{3/2}^{\mu\nu AB} = \frac{1}{4} \left( 3\eta^{\mu\nu} \epsilon^{AB} +2\Sigma^{\mu\nu AB}\right).
\ee
The projector property for each of these can be checked by using the algebra of $\Sigma$-matrices
\be
\Sigma^{\mu\alpha AE} \Sigma_{\alpha \, E}^{\,\,\,\nu \, B} = \frac{3}{4} \eta^{\mu\nu} \epsilon^{AB} - \Sigma^{\mu\nu AB},
\ee
which can be checked e.g. directly from the expression (\ref{Sigma}). We then write the one-form field $\chi_\mu^A$ as
\be\label{rho-dec}
\chi_\mu^A = \chi_\mu^{(1/2) A} +  \chi_\mu^{(3/2) A},
\ee
where $\chi^{(1/2)}_{\mu}{}^{A}=P_{1/2\, \mu \nu}^{AB}  \chi^{\nu}_{B} $ and $\chi^{(3/2)}_{\mu}{}^{A}=P_{3/2\, \mu \nu}^{AB}  \chi^{\nu}_{B} $.
Note that the spin 3/2 part satisfies
\be
\theta^{\mu AA'} \chi^{(3/2)}_{\mu A} = 0,
\ee
while the spin 1/2 part is of the form
\be
\chi^{(1/2)}_{\mu A} = \theta_{\mu AA'} \lambda^{A'}
\ee
for some two-component spinor $\lambda^{A'}$. We also note that $\Sigma$ viewed as an operator on the space spinor-valued one-forms on each irreducible representation acts as a multiplication operator, and thus
\be\label{S-rho}
\Sigma^{\mu\nu AB} \chi_{\nu B} = -\frac{3}{2} \chi^{(1/2)\mu A} + \frac{1}{2} \chi^{(3/2)\mu A} .
\ee

\subsection{Field equations}

Using the decomposition (\ref{rho-dec}) of $\chi_\mu^A$ into irreducible components we see that (\ref{RS-2}) implies that 
\be
\chi^{(1/2)}_{\mu A}=0.
\ee 
This means that
\be\label{RS-3}
\Sigma^{\mu\nu AB} \chi_{\nu B} =  \frac{1}{2} \chi^{\mu A},
\ee
and thus (\ref{RS-1}) implies that
\be\label{RS-4}
\partial^\mu \chi_{\mu A}=0.
\ee

We can use these implications of the field equations to see what equation the non-vanishing part of $\chi_\mu^A$ satisfies. Using (\ref{RS-3}) we can immediately rewrite the first equation in (\ref{RS-feqs}) as
\be
\chi^{\dagger\, \mu A'} = - \frac{\sqrt{2}}{m} \epsilon^{\mu\nu\rho\sigma} \theta_\nu^{AA'} \partial_\rho \chi_{\sigma A}.
\ee
We then substitute it into the second equation in (\ref{RS-feqs}), open up the product of two epsilon tensors, and use the first identity in (\ref{S-theta-id}). Taking into account the transversality (\ref{RS-4}) of $\chi_\mu^A$ we finally get
\be
(\partial^\alpha\partial_\alpha - m^2) \chi_\mu^A = 0,
\ee
which is the Klein-Gordon equation for the 4 out of 8 propagating components of $\chi_\mu^A$. It also identifies the parameter $m$ in the Lagrangian with the mass. We refrain from giving a second order description of the Rarita-Schwiner field, as it is rather cumbersome, unlike the case with Majorana and Weyl fermions.

\section{Spinor valued one-form description of a free Majorana fermion}
\label{sec:maj}

\subsection{Lagrangian}

Having fixed our spinor notations, and considered the 2-component formulations of usual Weyl and Dirac fermions, we are ready for the main objective of this paper, which is to study the Lagrangian stated in the Introduction. Let us write it keeping all the metrics involved explicitly
\be\label{L-m}
{\cal L}= 2 \epsilon_{AB} (\Sigma^{\mu\nu\,AC}\partial_\mu \rho_{\nu\,C}) (\Sigma^{\rho\sigma\,BD}\partial_\rho\rho_{\sigma\,D})   +\frac{3m^2}{2} \eta^{\mu\nu}\epsilon^{AB} \rho_{\mu\,A} \rho_{\nu\,B}.
\ee

We are now prepared to analyze what the field equations for (\ref{L-m}) imply. 

\subsection{Field equations}

The Euler-Lagrange equation for (\ref{L-m}) reads
\be\label{rho-feq}
 \Sigma^{\mu \nu AB} \partial_\nu (\Sigma^{\rho\sigma D}_{\,\,\,\, B}\partial_\rho\rho_{\sigma\,D})=\frac{3m^2}{4} \rho^{\mu A}.
\ee
Applying $\partial_\mu$ to this equation, and using the fact that the partial derivatives commute we immediately get
\be
\partial^\mu \rho_\mu^A = 0.
\ee
As in the case of the Rarita-Schwinger field, this equation is useful as a gauge-fixing condition helping to determine the propagating field content. Let us do this, and substitute the decomposition (\ref{rho-dec}) of the field $\rho_\mu^A$ into irreducible components. We see that the transverse part of one irreducible component determines that of the other:
\be
\partial^\mu\rho_\mu^{(1/2) A} +  \partial^\mu\rho_\mu^{(3/2) A} =0.
\ee
But then, using the fact (\ref{S-rho}) that the action of $\Sigma$ on $\rho$ is a multiple of the identity on each irreducible component, we can write (\ref{rho-feq}) as
\be\label{rho-feq-1}
- 2 \Sigma^{\mu \nu AB} \partial_\nu (\Sigma^{\rho\sigma\,D}_{\quad\! B}\partial_\rho\rho^{(1/2)}_{\sigma\,D})=\frac{3m^2}{4} \rho^{\mu A}.
\ee
We now note that the $\rho_{\mu A}^{(1/2)}$ irreducible component is of the form
\be
\rho_{\mu A}^{(1/2)} = \theta_{\mu AA'}\lambda^{A'}
\ee
for some spinor $\lambda^{A'}$. We can then project out of (\ref{rho-feq-1}) the spin $1/2$ component by multiplying this equation with $\theta_{\mu A}^{A'}$. Using
\be
\theta_{\mu A}^{A'} \Sigma^{\mu\nu AB} = \frac{3}{2} \theta^{\nu BA'},
\ee
as well as some elementary algebra of the soldering forms, we get
\be
(\partial^\mu \partial_\mu - m^2) \lambda^{A'} = 0
\ee
as a consequence of (\ref{rho-feq-1}). The spin $3/2$ component of this equation then determines the $\rho_{\mu A}^{(3/2)}$ part of $\rho_{\mu A}$ in terms of second derivatives of $\lambda^{A'}$. This shows that the theory (\ref{L-m}) is indeed about a massive propagating spin $1/2$ particle. We would now like to arrive at the same result via the method of Hamiltonian analysis, which clearly demonstrates what is going on. This will also allow us to treat a bit more general Lagrangian than (\ref{L-m}). 

\subsection{More general Lagrangian}

We now present a more general analysis, and consider instead the following Lagrangian:
\be\label{Lagr}
{\cal L}= 2 \epsilon_{AB} (\Sigma^{\mu\nu\,AC}\partial_\mu \rho_{\nu\,C}) (\Sigma^{\rho\sigma\,BD}\partial_\rho\rho_{\sigma\,D})   + \alpha \Sigma^{\mu\nu\,AB}\rho_{\mu\,A} \rho_{\nu\, B} + \beta \eta^{\mu\nu}\epsilon^{AB} \rho_{\mu\,A} \rho_{\nu\,B}.
\ee
When $\alpha=0$ and $\beta=3m^2/2$ we get the Lagrangian (\ref{L-m}). We could have repeated the above covariant analysis for this Lagrangian as well, but it becomes more messy. However, at the level of the Hamiltonian formulation there is no difficulty in adding the $\alpha$-term. 

Below we shall see that from the two "mass" terms seemingly present in (\ref{Lagr}), only a combination of the parameters turns out to have the meaning of mass. This can be seen from the fact (derived later) that when $\beta=0$ the theory (\ref{Lagr}) is topological with no propagating degrees of freedom. Thus, the parameter $\beta\not=0$ is essential for our construction, while $\alpha$ could be set to zero, as we have done in the considerations above. However, we decided to keep it to make the analysis more general, as the $\alpha$-term is a very natural one to add. Indeed, this is the mass term familiar from the Rarita-Schwinger Lagrangian (\ref{RS}). As a byproduct for $\beta=0$ we get what seems to be a new topological theory of fermions. 

\subsection{Projections of self-dual two-forms}

We now proceed with the Hamiltonian analysis of (\ref{Lagr}). For this we first need various projections of the two-forms $\Sigma^{AB}_{\mu\nu}$. First, it is easy to compute the temporal-spatial component of the two-forms $\Sigma^{AB}$. Using (\ref{Sigma}), we have
\be\label{Sigma-0i}
\Sigma_{0i}^{AB} \equiv \Sigma^{AB}_{\mu\nu} \left(\frac{\partial}{\partial t}\right)^\mu \left(\frac{\partial}{\partial x^i}\right)^\nu= \frac{1}{\sqrt{2}} m^i o^A o^B - \frac{1}{\sqrt{2}} \bar{m}^i \iota^A \iota^B - z^i o^{(A}\iota^{B)}=\frac{1}{\sqrt{2}} \sigma^{i AB}= \frac{\im}{2} T^{i\,AB},
\ee
where the objects $T^{i\,AB}$ were introduced in (\ref{T}), and $m^i, \bar m^i, z^i$  are the spatial components of the null tetrad ${l^\mu, n^\mu ,m^\mu, \bar m^\mu}$, see (\ref{ln}).

Now, using the easily derivable identities
\be
\im \epsilon^{ijk} z_i m_j = m_k, \qquad \im \epsilon^{ijk} z_i \bar{m}_j =-\bar{m}_k, \qquad \im \epsilon^{ijk} m_i \bar{m}_j = z_k,
\ee
we can easily compute
\be\label{Sigma-ij}
\frac{1}{2}\epsilon^{ijk} \Sigma_{ij}^{AB} = -\frac{\im}{\sqrt{2}} m^k o^A o^B + \frac{\im}{\sqrt{2}} \bar{m}^k \iota^A \iota^B + \im z^k o^{(A} o^{B)}= \frac{1}{2} T^{k\, AB}.
\ee
Thus, in particular we have
\be
\im \Sigma_{0i}^{AB} + \frac{1}{2} \epsilon_i{}^{jk}\Sigma_{jk}^{AB}=0,
\ee
which is the condition of self-duality with our conventions $\epsilon^{0123}=+1$.

\subsection{Projectors}

Before we write (\ref{Lagr}) in space plus time form, let us manipulate the combination that appears in the first "kinetic" term into a convenient form. We have
\be\label{S-d-rho}
\Sigma^{\mu\nu\,AC}\partial_\mu \rho_{\nu\,C}= - \Sigma_{0i}^{AC}(\partial_t \rho_{iC} - \partial_i \rho_{0C}) + \Sigma^{AC}_{ij} \partial_i \rho_{jC} \\ \nonumber = - \frac{\im}{2} \left( \partial_t(T^{iAC}\rho_{iC})-T^{iAC}\partial_i \rho_{0C} + \im \epsilon^{ijk} T^{kAC}\partial_i \rho_{jC}\right),
\ee
where we have used the expressions (\ref{Sigma-0i}), (\ref{Sigma-ij}). 

We now introduce to projectors
\be
P^{(3/2) ij AB}:= \frac{1}{3} \left( 2 \delta^{ij} \epsilon^{AB} + \epsilon^{ijk} T^{k AB}\right), \quad
P^{(1/2) ij AB}:= \frac{1}{3} \left( \delta^{ij} \epsilon^{AB} - \epsilon^{ijk} T^{k AB}\right).
\ee
These act on the space of objects of the type $\rho_{iA}$, and decompose it into two irreducible components - the spin $3/2$ and spin $1/2$ irreducible representations of the spatial rotation group. In writing a formula for the action, the natural contraction of unprimed spinors is used. 

We now decompose the spatial projection $\rho_{iA}$ of the original spinor valued one-form into its irreducible components $ \rho^{(3/2)}_{iA}$ and $ \rho^{(1/2)}_{iA}$. It is not hard to check that the spin $1/2$ component is of the form
\be
\rho^{(1/2)}_{iA} = -\frac{1}{3} T_{iA}{}^B \lambda_B
\ee
for some spinor $\lambda_B$. The prefactor is introduced so that $T^{iAB} \rho^{(1/2)}_{iB}=\lambda^A$. Thus, we write
\be
\rho_{iA} = \rho^{(3/2)}_{iA}  -\frac{1}{3} T_{iA}{}^B \lambda_B.
\ee
The two factors here are eigenvectors of the operator $\epsilon^{ijk} T^{kAB}$ of eigenvalues $+1$ and $-2$ respectively. Thus, we can write (\ref{S-d-rho}) as
\be
- \frac{\im}{2} \left( \partial_t \lambda^A -T^{iAC}\partial_i \rho_{0C} +\im \partial_i \rho^{(3/2)iA} +\frac{2\im}{3} T^{iAB} \partial_i \lambda_B \right).
\ee
From this we immediately see that only the $\lambda^A$ component of $\rho_\mu^A$ propagates, while all other fields are auxiliary. 

Let us also compute all other combinations that appear in the Lagrangian. We have
\be
\Sigma^{\mu\nu\,AB}\rho_{\mu\,A} \rho_{\nu\, B} = - 2\Sigma^{AB}_{0i} \rho_{0A}\rho_{iB} + \Sigma_{ij}^{AB} \rho_{iA}\rho_{jB}  =- \im T^{iAB} \rho_{0A}\rho_{iB} +\frac{1}{2} \epsilon^{ijk}T^{kAB} \rho_{iA}\rho_{jB}.
\ee
The first term here contains just the spin $1/2$ component $\lambda^A$. The second term can be computed again using the fact that the $\epsilon^{ijk}T^{kAB}$ operator takes specific values on the two irreducible components. We get, overall
\be
\Sigma^{\mu\nu\,AB}\rho_{\mu\,A} \rho_{\nu\, B} = \im \rho_{0}^{A} \lambda_A + \frac{1}{3}\lambda^A \lambda_A +\frac{1}{2} \rho^{(3/2)}_{iA} \rho^{(3/2) iA}.
\ee

We now compute the last term
\be
g^{\mu\nu}\epsilon^{AB} \rho_{\mu A} \rho_{\nu B} = \rho_0^A \rho_{0 A} + \rho_{iA} \rho^{iA}= \rho_0^A \rho_{0 A} - \frac{1}{3}\lambda^A \lambda_A + \rho^{(3/2)}_{iA} \rho^{(3/2) iA}.
\ee

\subsection{Hamiltonian analysis}

We first write the space plus time decomposition of the full Lagrangian
\be
{\cal L}=  \frac{1}{2} \left( \partial_t \lambda^A -T^{iAC}\partial_i \rho_{0C} +\im \partial_i \rho^{(3/2)iA} +\frac{2\im}{3} T^{iAB} \partial_i \lambda_B \right)^2\\ \nonumber + \alpha\left(\im \rho_{0}^{A} \lambda_A + \frac{1}{3}\lambda^A \lambda_A +\frac{1}{2} \rho^{(3/2)}_{iA} \rho^{(3/2) iA}\right)  +\beta\left( \rho_0^A \rho_{0 A} - \frac{1}{3}\lambda^A \lambda_A + \rho^{(3/2)}_{iA} \rho^{(3/2) iA}\right).
\ee
The Hamiltonian analysis is now easy. First, the momentum conjugate to $\lambda_A$ is
\be
\pi^A = \partial_t \lambda^A -T^{iAC}\partial_i \rho_{0C} +\im \partial_i \rho^{(3/2)iA} +\frac{2\im}{3} T^{iAB} \partial_i \lambda_B.
\ee
The Hamiltonian is
\be
{\cal H} = \frac{1}{2} \pi^A \pi_A + \pi^A \left( T_A^{iC}\partial_i \rho_{0C} -\im \partial_i \rho_A^{(3/2)i} -\frac{2\im}{3} T_A^{iB} \partial_i \lambda_B \right) \\ \nonumber
- \alpha\left(\im \rho_{0}^{A} \lambda_A + \frac{1}{3}\lambda^A \lambda_A+\frac{1}{2} \rho^{(3/2)}_{iA} \rho^{(3/2) iA}\right)  - \beta\left( \rho_0^A \rho_{0 A} - \frac{1}{3}\lambda^A \lambda_A + \rho^{(3/2)}_{iA} \rho^{(3/2) iA}\right).
\ee
Now the spin $3/2$ field $\rho^{(3/2)}_{iA}$ can be eliminated from the action by solving its field equation. We have
\be\label{rho-i}
(2\beta+\alpha)\rho^{(3/2)}_{iA} = - \im P^{(3/2)}(\partial_i \pi_A),
\ee
where the projection on the spin $3/2$ component is taken. This can be solved when $2\beta\not=\alpha$. Substituting this solution back we get a (partially) reduced Hamiltonian
\be
{\cal H} = \frac{1}{2} \pi^A \pi_A + \pi^A \left( T_A^{iC}\partial_i \rho_{0C} -\frac{2\im}{3} T_A^{iB} \partial_i \lambda_B \right) - \frac{P^{(3/2)}(\partial_i \pi_A) \partial^i \pi^A}{2(2\beta+\alpha)} \\ \nonumber
- \alpha\left(\im \rho_{0}^{A} \lambda_A + \frac{1}{3}\lambda^A \lambda_A \right)  - \beta\left( \rho_0^A \rho_{0 A} - \frac{1}{3}\lambda^A \lambda_A \right).
\ee
We also see now that when $\beta=0$ the field $\rho_0^A$ plays the role of a Lagrange multiplier for a constraint. This constraint generates gauge transformations on the phase space $\lambda^A,\pi_A$, and completely kills all propagating degrees of freedom. This does not happen for a non-zero $\beta$. In this case the field $\rho_0^A$ can also be eliminated using its field equation. We get
\be\label{rho-0}
\rho_0^A = \frac{1}{2\beta}\left( T^{iAB}\partial_i \pi_B -\im \alpha \lambda^A\right).
\ee
Substituting this back we find a fully reduced Hamiltonian
\be\label{H-1}
{\cal H} = \frac{1}{2} \pi^A \pi_A + \frac{2\im}{3} \pi_A  T^{iAB} \partial_i \lambda_B  - \frac{P^{(3/2)}(\partial_i \pi_A) \partial^i \pi^A}{2(2\beta+\alpha)} \\ \nonumber
+ \frac{(\beta-\alpha)}{3}\lambda^A \lambda_A + \frac{1}{4\beta} \left( T^{iAB}\partial_i \pi_B -\im \alpha \lambda^A\right)\left( T_A^{jC}\partial_j \pi_C -\im \alpha \lambda_A\right).
\ee
We now put similar terms together to obtain
\begin{align}\label{H-Maj'}
\mathcal H=\frac{1}{2} \pi^A  \pi_A- \im \left(\frac{4\beta +3 \alpha}{6\beta} \right) \pi^A T^{i \phantom A B }_{\phantom i A} \partial_i \lambda_B +& \left( \frac{(2\beta-3\alpha)(2\beta+\alpha)}{12\beta} \right) \lambda^A \lambda_A \notag \\
+& \left( \frac{10\beta+3\alpha}{12\beta(2\beta+\alpha)}  \right)  (\partial^i \pi^A) (\partial_i \pi_A)  .
\end{align}
Here we have used the explicit form of the $P^{(3/2)}$ projector, and using (\ref{T-algebra}) expanded the product of two $T$Õs in the last term in (\ref{H-1}). We have also dropped, after integrating by parts, the terms containing the $\epsilon^{ijk}$ tensor. Let us now define the parameters $C_1$ and $C_2$ in terms of $\alpha$ and $\beta$ as
\begin{align}
C_1=&\frac{4\beta+3\alpha}{6\beta} \, ,  &     C_2=2\beta+\alpha \, .
\end{align}
Then, the Hamiltonian (\ref{H-Maj'}) can be rewritten as 
\begin{equation}\label{H-Maj}
\mathcal H=\frac{1}{2} \pi^A  \pi_A- \im C_1 \pi^A T^{i \phantom A B}_{\phantom i A} \partial_i \lambda_B +  \frac{(1-C_1)C_2}{2}  \lambda^A \lambda_A+ \frac{(1+C_1)}{2C_2}  (\partial^i \pi^A) (\partial_i \pi_A)  .
\end{equation}

\subsection{Evolution equations}

We would now like to see what dynamics the Hamiltonian (\ref{H-Maj}) gives rise to for $\lambda^A$. The reduced Hamiltonian equation for $\lambda_A$ is
\begin{align}\label{ham-eq-lambda}
\partial_t \lambda_A=&\left(  \frac{\overrightarrow \partial}{\partial \pi^A}   \mathcal H \right)\, ,\notag\\
\partial_t \lambda_A=& \pi_A-\im C_1 T^{i\phantom A B}_{\phantom i A} \partial_i \lambda_B -\frac{(1+C_1)}{C_2} \, \Delta \pi_A \, ,
\end{align}
where $\Delta=\partial_i\partial^i$ is the Laplacian. The Hamiltonian equation of $\pi^A$ is
\begin{align}\label{ham-eq-pi}
\partial_t \pi^A=&-\left( \mathcal H \frac{\overleftarrow \partial}{\partial \lambda_A}   \right)\, ,\notag\\
\partial_t \pi^A=& \im C_1 T^{iA B} \partial_i \pi_B -(1-C_1)C_2 \,  \lambda^A \, .
\end{align}
Differentiating (\ref{ham-eq-lambda}) with respect to time, we get
\begin{equation}\label{der-tt-lambda}
\partial_t \partial_t \lambda_A=\partial_t \pi_A  -\im C_1 T^{i \phantom A B}_{\phantom i A} \,\partial_0 \partial_i\lambda_B-\frac{(1+C_1)}{C_2}\, \partial_t \Delta \pi_A \, .
\end{equation} 
Applying the operator $-\im C_1 T^{i \phantom A B}_{\phantom i A} \partial_i$ to (\ref{ham-eq-lambda}), we obtain
\begin{equation}
-\im C_1 T^{i \phantom A B}_{\phantom i A} \partial_t \partial_i \lambda_B=-\im C_1 T^{i \phantom A B}_{\phantom i A}\, \partial_i \pi_B+C_1^2\, \Delta \lambda_A+\im \frac{C_1(1+C_1)}{C_2} T^{i \phantom A B}_{\phantom i A} \,\partial_i \Delta \pi_B \, ,
\end{equation}
where we have use product of two $T$'s. Applying the operator $-(1+C_1)/C_2 \, \Delta$ to (\ref{ham-eq-pi}), we find
\begin{equation}
-\frac{(1+C_1)}{C_2}\,  \partial_t \Delta\pi_A=-\im \frac{C_1(1+C_1)}{C_2} T^{i \phantom A B}_{\phantom i A} \, \partial_i \Delta \pi_B+ (1-C_1^2)\, \Delta \lambda_A \, .
\end{equation}
Adding the two equation above and (\ref{ham-eq-pi}), we get in (\ref{der-tt-lambda})
\begin{equation}
(\partial_t \partial_t- \Delta+ m^2) \lambda_A=0 \, ,
\end{equation}
where 
\begin{equation}
m^2\equiv (1-C_1) C_2= \frac{(2\beta-3\alpha)(2\beta+\alpha)}{6\beta} \, ,
\end{equation}
that is, twice the coefficient of the $\lambda \lambda$ term in the Hamiltonian. Thus, the evolution equation resulting from (\ref{H-Maj}) is the usual Klein-Gordon equation for each component of the spinor $\lambda^A$. 

We note that for $\alpha=0$ this simplifies to $m^2=2\beta/3$. Thus, the term proportional to $\alpha$ in our starting Lagrangian (\ref{Lagr}) is not essential to get the dynamics of a massive Majorana fermion. We have included this term for completeness, because it is quite natural to add to the Lagrangian. We also see that the $\beta$-term is essential for a non-trivial dynamics, because the limit $\beta\to 0$ gives infinite mass, and thus effectively removes the propagating degree of freedom that we are describing. It can be seen from the above Hamiltonian analysis that the Lagrangian (\ref{Lagr}) with $\beta=0$ is in fact a topological theory with no propagating degrees of freedom. Indeed, when $\beta=0$ the theory has extra symmetry, generated by the constraint that gets imposed by varying with respect to $\rho_0^A$ that in this case receives an interpretation of the Lagrange multiplier. When $\beta\not=0$ this "topological" symmetry is broken, and one gets a propagating spin $1/2$ mode. 

\subsection{Field redefinition}

We now show that the Hamiltonian (\ref{H-Maj}) is just the original Majorana fermion Hamiltonian (\ref{H-Maj-orig}) in disguise. Thus, let us consider a field redefinition:
\be\label{l-shift}
\lambda^A \to \lambda^A + \im\gamma T^{i\, AB} \partial_i \pi_B. 
\ee
It is not hard to see that this is a canonical transformation that leaves the $\pi^A \partial_t \lambda_A$ presymplectic one-form intact. Therefore, substituting a shifted field into (\ref{H-Maj}) we can choose the coefficient $\gamma$ so that all the $(\partial^i \pi^A)(\partial_i \pi_A)$ terms cancel. One can show that
\be
\gamma = \frac{1+C_1}{(1-C_1)C_2}
\ee
does the job and that the resulting Hamiltonian for the shifted field $\lambda^A$ is
\be\label{H-redef}
{\cal H} = \frac{1}{2} \pi^A  \pi_A+ \im \pi^A T^{i \phantom A B}_{\phantom i A} \partial_i \lambda_B +  \frac{m^2}{2}  \lambda^A \lambda_A.
\ee
We note that one of the solutions of a quadratic equation for $\gamma$ was chosen, with the other just giving an opposite sign in front of the the $\im \pi^A T^{i \phantom A B}_{\phantom i A} \partial_i \lambda_B$ term in the Hamiltonian. The above Hamiltonian is of course exactly the usual Majorana Hamiltonian (\ref{H-Maj-orig}), after an additional simple rescaling of the fields $\lambda^A$ and $\pi^A$ that puts a factor of $m$ in front of the $\pi^A \pi_A$ term.  

Thus, we have shown that the Lagrangian (\ref{Lagr}), after all auxiliary non-propagating fields are eliminated, and after a simple shift of $\lambda^A$, leads to precisely the same Hamiltonian description of a 2-component spinor $\lambda^A$ as the original first-order Lagrangian (\ref{majorana}). This finishes our description of a single massive Majorana particle in terms of a spinor-valued one-form.

\subsection{Reality condition}

From our discussion of second-order formulation of fermions above we know that such a formulation must be supplemented with a reality condition. In the usual case this is the Dirac equation relating a primed spinor to the derivative of an unprimed one. In our case the appropriate reality condition can be worked out starting from the reduced Hamiltonian in the form (\ref{H-redef}). Indeed, we know that the relevant reality condition at this level is simply $\pi^A = \im \, m (\lambda^\star)^A$, see (\ref{pi-weyl}). Using the shift (\ref{l-shift}), this can be translated as a condition on the original spinor field. One gets:
\be\label{reality}
\im\, m (\lambda^\star)^A = \pi^A +\im(1+C_1) T^{i\,AB}\partial_i \lambda_B - \frac{(1+C_1)^2}{m^2} \Delta \pi^A.
\ee
This reality condition guarantees that the reduced Hamiltonian (\ref{H-Maj}) is Hermitian. It is also sufficient for the purposes of determining the mode decomposition of the field $\lambda^A$ in terms of creation-annihilation operators. The components of the original spinor-valued one-form field $\rho_\mu^A$ can then be determined in terms of $\pi^A, \lambda^A$ via (\ref{rho-i}), (\ref{rho-0}). This gives everything that is necessary for the decomposition of $\rho_\mu^A$ into modes. We will not give the corresponding expressions as we do not need them in this paper. One could also write the reality condition in spacetime form, as a condition directly on the original field $\rho_{\mu A}$. As in the case of the usual Majorana theory, this is a differential condition, where the complex conjugate is related to a derivative of the original field.


\section{Spinor valued one-form description of a Dirac fermion}
\label{sec:dirac}

In this section we generalize the above one-form description of a Majorana fermion to the case of a Dirac fermion. We only consider the free theory, postponing the analysis of possible interactions (in particular with the electromagnetic field) to later work. We are brief in this section, as it exactly parallels the above treatment. 

\subsection{Lagrangian}

Here we will follow the same recipe that was used in the construction of the Dirac fermion Lagrangian from two uncoupled 2-component Majorana Lagrangians of equal mass, see subsection 3.5.  Thus, the spinor-valued one-form Lagrangian of two uncoupled Majorana fermions of equal mass  is given by
\begin{align}\label{Dirac-lagrangian}
\mathcal L_{\rm D}=& 2\epsilon_{AB} \left[ \left(  \Sigma^{\mu\nu\, AC} \partial_\mu \rho^{\bf (1)}_{\nu C} \right)\left(  \Sigma^{\lambda \sigma\, BD} \partial_\lambda \rho^{\bf (1)}_{\sigma D} \right) + \left(  \Sigma^{\mu\nu\, AC} \partial_\mu \rho^{\bf (2)}_{\nu C} \right)\left(  \Sigma^{\lambda \sigma\, BD} \partial_\lambda \rho^{\bf (2)}_{\sigma D} \right)  \right]\notag \\
&+\left( \alpha\, \Sigma^{\mu\nu\, AB} + \beta \eta^{\mu\nu} \epsilon^{AB}   \right)\left( \rho^{\bf (1)}_{ \mu A}  \rho^{\bf (1)}_{\nu B} + \rho^{\bf (2)}_{ \mu A}  \rho^{\bf (2)}_{\nu B} \right)     \, ,
\end{align}
where the boldface upper indices in parenthesis ${\bf (1)}$ and ${\bf (2)}$ label the two uncoupled spinor-valued one-form fields. Now, making the following complex field transformation
\begin{align}\label{complex-transformation}
\rho^{\bf (1)}_{\mu\, A}=&\frac{1}{\sqrt{2}} \left( \omega_{\mu\, A} +\upsilon_{\mu \, A} \right)  \, , \notag \\
\rho^{\bf (2)}_{\mu\, A}=&\frac{\im}{\sqrt{2}} \left( \omega_{\mu\, A} -\upsilon_{\mu\, A} \right)\, ,
\end{align} 
we get
\begin{align}\label{Dirac-after-trans}
\mathcal L_{\rm D}= 4\epsilon_{AB} \left(  \Sigma^{\mu\nu\, AC} \partial_\mu \omega_{\nu C} \right) \left(  \Sigma^{\lambda \sigma\, BD} \partial_\lambda \upsilon_{\sigma D} \right)
+2\left( \alpha\, \Sigma^{\mu\nu\, AB} + \beta \eta^{\mu\nu} \epsilon^{AB}   \right)\,  \omega_{\mu A} \upsilon_{\nu B}   \, .
\end{align}
It is obvious that this Lagrangian is invariant under the global $U(1)$ symmetry
\begin{align}
\omega_{\mu} &\to e^{-\im \varphi}\, \omega_{\mu}   \, ,    &  \upsilon_{\mu} &\to e^{\im \varphi}\, \upsilon_{\mu}  \, .
\end{align}
This is our spinor-valued one-form version of the second-order Dirac Lagrangian (\ref{Dirac-2-order}).

\subsection{Hamiltonian analysis}

We now perform the space plus time split. The relevant formulas are
\be
\Sigma^{\mu\nu\, AC} \partial_\mu \omega_{\nu C}  = - \frac{\im}{2} \left( \partial_t \xi^A +\im A_0 \xi_A -T^{iAC}\partial_i \omega_{0C} +\im \partial_i \omega^{(3/2)iA} +\frac{2\im}{3} T^{iAB} \partial_i \xi_B \right),
\ee
where
\be
\omega^{iA} = \omega^{(3/2)iA} - \frac{1}{3} T_{iA}{}^B \xi_B,
\ee
and thus $\xi^A = T^{iAB}\omega_{iB}$. For the field $\upsilon_\mu^A$ one obtains a similar expression, with the exception of a different sign in front of the connection. We define $\chi^A = T^{iAB}\upsilon_{iB}$. The other quantities that appear in the Lagrangian are expanded as follows. 
\be
\Sigma^{\mu\nu\,AB}\omega_{\mu\,A} \upsilon_{\nu\, B} = \frac{\im}{2} \omega_{0}^{A} \chi_A +\frac{\im}{2} \upsilon_{0}^A \xi_A + \frac{1}{3}\xi^A \chi_A +\frac{1}{2} \omega^{(3/2)}_{iA} \upsilon^{(3/2) iA}.
\ee
The metric containing term expands to
\be
g^{\mu\nu}\epsilon^{AB} \omega_{\mu A} \upsilon_{\nu B} = \omega_0^A \upsilon_{0 A}  - \frac{1}{3}\xi^A \chi_A + \omega^{(3/2)}_{iA} \upsilon^{(3/2) iA}.
\ee
The momenta conjugate to $\xi_A, \chi_A$ are, respectively
\be
\pi^A = \partial_t \chi^A - \im A_0 \chi_A -T^{iAC}\partial_i \upsilon_{0C} +\im \partial_i \upsilon^{(3/2)iA} +\frac{2\im}{3} T^{iAB} \partial_i \chi_B, \\ \nonumber
\eta^A = \partial_t \xi^A +\im A_0 \xi_A -T^{iAC}\partial_i \omega_{0C} +\im \partial_i \omega^{(3/2)iA} +\frac{2\im}{3} T^{iAB} \partial_i \xi_B.
\ee
The Hamiltonian is
\be
{\cal H} = \pi^A \eta_A - \im A_0 (\pi^A \xi_A - \eta^A \chi_A) + \pi^A \left( T_A^{iC}\partial_i \omega_{0C} -\im \partial_i\omega_A^{(3/2)i} -\frac{2\im}{3} T_A^{iB} \partial_i \xi_B \right) \\ \nonumber 
+ \eta^A \left( T_A^{iC}\partial_i \upsilon_{0C} -\im \partial_i\upsilon_A^{(3/2)i} -\frac{2\im}{3} T_A^{iB}\partial_i \chi_B \right) 
\\ \nonumber
- \alpha\left(\im \omega_{0}^{A} \chi_A +\im \upsilon_{0}^A \xi_A + \frac{2}{3}\xi^A \chi_A  +\omega^{(3/2)}_{iA} \upsilon^{(3/2) iA}\right)  - 2\beta\left( \omega_0^A \upsilon_{0 A}  - \frac{1}{3}\xi^A \chi_A + \omega^{(3/2)}_{iA} \upsilon^{(3/2) iA}\right).
\ee
As in the Majorana case, we now eliminate the non-propagating modes. We have
\be
(2\beta+\alpha)\omega^{(3/2)}_{iA} = - \im P^{(3/2)}(\partial_i \eta_A), \qquad (2\beta+\alpha)\upsilon^{(3/2)}_{iA} = - \im P^{(3/2)}(\partial_i \pi_A), \\ \nonumber
\omega_0^A = \frac{1}{2\beta}\left( T^{iAB} \partial_i \eta_B -\im \alpha \xi^A\right), \qquad
\upsilon_0^A = \frac{1}{2\beta}\left( T^{iAB} \partial_i \pi_B -\im \alpha \chi^A\right).
\ee
We now substitute this back into the Hamiltonian, and obtain the fully reduced Hamiltonian in the form
\begin{align}
\mathcal H =& \pi^A \eta_A - \im A_0 (\pi^A \xi_A - \eta^A \chi_A)- \im C_1\left(\pi^A T^{i \phantom A B}_{\phantom i A} \partial_i \xi_B +\eta^A T^{i \phantom A B}_{\phantom i A} \partial_i \chi_B    \right)\notag \\ &+  m^2 \, \chi^A \, \xi_A + \frac{(1+C_1)}{C_2} \,   (\partial^i \pi^A) (\partial_i \eta_A)  + \frac{C_1}{C_2} \epsilon^{ijk} T^{k AB} \partial_i \pi_A \partial_j \eta_B \, .
\end{align}
This Hamiltonian is invariant under the global $U(1)$ transformation
\begin{align}\label{dirac-ham}
\pi  &\to e^{-\im \varphi}\pi   \, ,    &  \xi &\to e^{\im \varphi}\, \xi  \, , \notag \\
\eta  &\to e^{\im \varphi}\eta   \, ,    &  \chi &\to e^{-\im \varphi}\, \chi \, .
\end{align}
Here, unlike in (\ref{H-Maj}) we kept the last, anti-symmetric in derivatives term. It vanishes upon integration by parts when the derivatives commute, but will not vanish once the interaction with the electromagnetic field is switched on. We do not study this in the present work.

\subsection{Field redefinition}

We can now do a similar field redefinition as in the Majorana fermion case. Thus, we perform the following shifts
\be
\xi^A \to \xi^A +\im\frac{1+C_1}{(1-C_1)C_2} T^{i\, AB} \partial_i \eta_B, \qquad
\chi^A \to \chi^A +\im\frac{1+C_1}{(1-C_1)C_2} T^{i\, AB} \partial_i \pi_B.
\ee
The Hamiltonian then takes the form
\be
\mathcal H_{\rm D}=\pi^A\eta_A- \im A_0 (\pi^A \xi_A - \eta^A \chi_A)+\im \pi^A T^{i \phantom A B}_{\phantom i A} \partial_i \xi_B +\im \eta^A T^{i \phantom A B}_{\phantom i A} \partial_i \chi_B  +  m^2\, \chi^A \, \xi_A,
\ee
which is the usual (free) Dirac Hamiltonian.  

\section{Discussion}

The main result of this paper is a description (\ref{intr-L}) of a single massive uncharged spin $1/2$ particle (Majorana fermion) using a spinor-valued one-form field $\rho_\mu^A$. This is a greatly redundant description, as is clear from the fact that from $4\times 2=8$ components of the field $\rho_\mu^A$ only 2 components propagate. The other 6 components are auxiliary fields that are eliminated when the corresponding second class constraints present in the system are solved for. After this is done one gets, after a simple local field redefinition (\ref{l-shift}) of the fermion field, the standard Majorana fermion Hamiltonian (\ref{H-redef}). This description is easy to generalize to the case of a (free) Dirac fermion (i.e. an electrically charged spin $1/2$ particle with its anti-particle), simply by considering a pair of initially uncoupled Majorana fermions of the same mass and then performing a complex rotation that exhibits the ${\rm U}(1)$ symmetry. 

As we have seen from the Hamiltonian analysis, when $\beta=0$ the Lagrangian (\ref{Lagr}) describes a topological theory without any propagating degrees of freedom. As an aside remark let us point out that this topological Lagrangian can be written without any mentioning of the metric by "integrating in" an extra spinor field $\lambda^A$. We can then write 
\be\label{L-top}
{\cal L}_{\rm top} = \lambda_A \Sigma^{AB}\wedge d\rho_B + \alpha \Sigma^{AB}\wedge \rho_A \wedge \rho_B + (\Sigma^{CD}\wedge \Sigma_{CD}) \lambda^A \lambda_A.
\ee
Putting the coefficient in front of the last term to be unity is without loss of generality, for this can always be achieved by rescaling of $\rho,\lambda$. When $\Sigma^{AB}$ are the self-dual two-forms (\ref{Sigma}) for the Minkowski metric, integrating out the field $\lambda^A$, one gets back the topological $\beta=0$ version of the Lagrangian (\ref{Lagr}). But of course (\ref{L-top}) makes sense for arbitrary $\Sigma^{AB}$, not necessarily corresponding to Minkowski metric, and not even necessarily satisfying the "metricity" equation $\Sigma^{AB}\wedge \Sigma^{CD}\sim \epsilon^{A(C}\epsilon^{D)B}$. The Lagrangian (\ref{L-top}) could be interesting in its own right as a simple topological theory of fermions coupled to gravity (via $\Sigma^{AB}$). The Hamiltonian analysis of this topological theory that is a subcase of the more general analysis presented in the main text is a side result of the present work.

The above first-order in derivatives form is also possible for the full $\beta\not=0$ Lagrangian, but in this case one needs to add to (\ref{L-top}) a term that explicitly contains the metric, i.e. the $\beta$-term of (\ref{Lagr}). It is thus clear that a second order in derivatives nature of the Lagrangian we used is not essential. If desired, one can always integrate in an extra field to make it first order. We have decided to work with the second order version because it is less redundant, as containing just a single field $\rho_\mu^A$. 

The above remark relating our Lagrangian (\ref{Lagr}) to one of a topological theory suggests an intriguing way to think about our construction. Indeed, we have taken a topological theory and added to it the $\beta$-term that breaks the topological symmetry and thus introduces propagating degrees of freedom. The very same phenomenon occurs in the Plebanski formulation of general relativity (GR), see e.g. \cite{Krasnov:2009pu} for a description. In this formulation GR appears when one adds to the topological BF theory Lagrangian a Lagrange multiplier term that breaks the topological symmetry. A more general way to break this symmetry is to add to the BF Lagrangian a potential for the B field, as is studied in e.g. \cite{Krasnov:2008fm}. What happens in our description of fermions is quite similar, and, in fact, this analogy was what guided us to the Lagrangian presented in this paper. 

It is also useful to compare our description to other constructions available in the literature. One such attempt closest to us in motivations was given in \cite{Capovilla:1991qb}. This work proposed an action principle with the kinetic term being essentially our first term in (\ref{L-top}), and then a Lagrange multiplier term added to eliminate the unwanted components of the one-form field $\rho_\mu^A$. However, the constraint $\Sigma^{(AB}\wedge \rho^{C)}=0$ resulting when the Lagrange multipliers are varied contains $4\times 4$ equations. It is clear that not all of these equations can be those on the one-form field $\rho_\mu^A$ with its $2\times 4$ components. Thus, some of these equations are those on $\Sigma^{AB}$. It can then be show that there is a non-trivial $\rho_\mu^A\not=0$ solution to the constraint only when $\Sigma^{AB}$ satisfy their simplicity constraint $\Sigma^{(AB}\wedge\Sigma^{CD)}=0$. However, for our purposes of extending the fermionic coupling to the class of theories in \cite{Krasnov:2008fm} this is unsatisfactory. This inability of the existing formulations to deal with more general two-forms is one of the motivations behind the construction in this paper. Thus, our Lagrangian (\ref{intr-L}) does not assume any condition on the $\Sigma^{AB}$ two-form field. But it explicitly uses the metric, unlike the Lagrangian in \cite{Capovilla:1991qb}. This is not a cause of concern from our standpoint, as Lagrangians of the type studied here {\it can} arise once a general diffeomorphism invariant gauge theory is expanded around an appropriate background, see \cite{Krasnov:2011hi}. 

Our other comment is about a much more trivial way to obtain a description of fermions by one-form valued fields. Indeed, one can just take the second-order Majorana Lagrangian (\ref{weyl-chiral}) and replace in it every occurrence of the field $\lambda^A$ with $\theta^{\mu AA'} \rho_{\mu A'}$, where $\rho_{\mu A'}$ is a spinor and Grassmann-valued one-form. It is clear that the resulting $\rho_{\mu A'}$ Lagrangian will continue to propagate the spin 1/2 particle. It is worth emphasizing that this is not what has been done in this work. Instead, we studied a Lagrangian of the type that is known to arise when expanding a diffeomorphism invariant gauge theory around an appropriate background, as in \cite{Krasnov:2011hi}. 

Let us now discuss the open problems related to our construction. The first and foremost is that of coupling to other fields. Indeed, we have motivated our one-form based description of fermions by the idea to put this one-form together with the gauge fields for gravity and Yang-Mills into a large super-connection. Interactions should then be obtained by expanding the basic action around an appropriate background. This has not been realized in the present paper, as we have considered a free theory. The main reason for this is that any such discussion would require introducing a rather heavy machinery of Lie super-algebras. Thus, we have decided to postpone such studies to future publications. 

The other important open problem that our construction has to face is that of reality conditions, or, equivalently, the issue of unitarity. On one hand, having the explicit expression (\ref{l-shift}) for the field redefinition to the usual Majorana Lagrangian variables, we can state the reality conditions as in (\ref{reality}). After this is done, we get an equivalent description of the  Majorana fermion. On the other hand, it is clearly necessary to understand the reality as some condition on the basic one-form field $\rho_\mu^A$. We have not attempted to find this here because the issues of reality are likely to be tied with the issues of gauge field and gravity couplings. Indeed, if at all possible, it will most likely be that the reality conditions for the fermions can only be understood together with these for the other fields, and this is an open problem even in (the gauge-theoretic description of) the gravity sector. So, this whole set of open questions remains the subject of future work.

\section*{Acknowledgements} ATG was partially supported by Fondecyt grant 3130333. KK was supported by an ERC Starting Grant 277570-DIGT, as well as partially by the Alexander von Humboldt foundation, Germany. 

\appendix

\section{Appendix: Two-component spinors}
\label{sec:prelim}

In this section we remind the reader how the Weyl and Dirac fermions are described using 2-component spinors. Such a description is now a part of at least some quantum field theory treatments, see e.g. \cite{Srednicki:2007qs}. However, unlike the standard in the particle theory literature notation of dotted and undotted spinors, we use the notation familiar from the GR literature. Here we give a brief description of fermions using this language. 

\subsection{${\rm SL}(2,\C)$ spinors}

A note is in order about our spinor conventions. We have two types of spinors, those with unprimed indices $A,B,\ldots$, which constitute the fundamental representation of ${\rm SL}(2,\C)$, and those with primed indices $A',B',\ldots$, which form the complex conjugate representation. Unless otherwise noted, all our two-component spinors are Grassmann-valued objects. Taking a Hermitian conjugation of an unprimed spinor one obtains a primed spinor:
\be
(\lambda^A)^\dagger = (\lambda^\dagger)^{A'}.
\ee

The object $\epsilon_{AB}=\epsilon_{[AB]}$ is the anti-symmetric rank 2 spinor providing an isomorphism between unprimed spinors and their duals (i.e. an isomorphism between spinors with upper and lower indices). The inverse of $\epsilon_{AB}$ is defined via $\epsilon^{AB}\epsilon_{AC}=\delta_C{}^B$, where $\delta_A{}^B$ is the Kronecker delta. Sometimes we also write $\epsilon_A{}^B=\delta_A{}^B$. The raising and lowering of indices is according to:
\be
\lambda^A=\epsilon^{AB}\lambda_B, \qquad \lambda_B = \lambda^A \epsilon_{AB}.
\ee
In other words, the rule is that the spinor indices to be contracted are always located up to down if one reads the formula from the left. The raising and lowering of primed spinor indices is defined similarly with the help of $\epsilon_{A'B'}$ and $\epsilon^{A'B'}$ anti-symmetric tensors. Note that we do not put a bar above the epsilon. 

As is usual in the 2-component spinor literature, we shall often use an index-free notation:
\be
\lambda^A \xi_A := \lambda \xi, \qquad (\lambda^\dagger)_{A'} (\xi^\dagger)^{A'} = \lambda^\dagger \xi^\dagger.
\ee
This is a natural convention, for we have
\be
(\lambda\xi)^\dagger = \xi^\dagger \lambda^\dagger.
\ee

In the spinor formalism the metric is described by a soldering form (the spinor analog of a tetrad), which is a one-form with values in the rank 2 mixed spinors: $\theta_\mu^{AA'}$. For real metrics our convention is that the soldering form is Hermitian $\bar{\theta}^{AA'}=\theta^{AA'}$, which is the standard reality condition in the particle theory literature, e.g. \cite{Srednicki:2007qs}. The soldering form provides an isomorphism between the tangent space to the spacetime manifold and the space of rank 2 mixed spinors. The spacetime metric is obtained as 
\be\label{metric}
\eta_{\mu\nu}=- \theta_\mu^{AA'}\theta_\nu^{BB'}\epsilon_{AB}\epsilon_{A'B'},
\ee
where the minus sign is necessary to obtain a metric of signature $(-,+,+,+)$.

\subsection{A null tetrad}

When working with spinors, it proves to be very convenient to introduce a tetrad all 4 components of which are null. It consists of 2 real null one-forms $l_\mu,n_\mu$ and two complex-conjugate null one forms $m_\mu,\bar{m}_\mu$. These satisfy the following conditions \begin{align}
l^{\mu} n_{\mu}&=-1 \; , & m^{\mu} \overline{m}_{\mu}&=1  \; ,
\end{align} 
with all the other contractions equal to zero. Moreover, its relation with the Minkowski tetrad $\{t,x,y,z\}$ is
\begin{align}\label{ln}
l^{\mu}=& \frac{1}{\sqrt{2}} \left( t^{\mu}+z^{\mu} \right) \;\; , &  n^{\mu}=& \frac{1}{\sqrt{2}} \left( t^{\mu}-z^{\mu} \right) \;\; ,\\
m^{\mu}=& \frac{1}{\sqrt{2}} \left( x^{\mu}+i y^{\mu} \right) \;\; , &  \overline{m}^{\mu}=& \frac{1}{\sqrt{2}} \left( x^{\mu}-i y^{\mu} \right) \, .
\end{align}
One also introduces a basis in the space of primed and unprimed spinors. The basis spinors are denoted by $\iota^A, o^A$ for unprimed spinors and $\iota^{A'}, o^{A'}$ for primed (to avoid the clatter of notations we use $\bar{\iota}^{A'}:=\iota^{A'}, \bar{o}^{A'}:=o^{A'}$). Note that the basis spinors $\iota^A, o^A$ are the usual c-valued spinors, not Grassmann-valued. Our normalisation convention is:
\be
\iota^A o_A = 1,
\ee
and similarly for the primed basis spinors. The tetrad $\theta_\mu^{AA'}$ can be expanded in the basis spinors as follows:
\be\label{theta-io}
\theta_\mu^{AA'} = l_\mu o^A o^{A'} + n_\mu \iota^A \iota^{A'} + m_\mu o^A \iota^{A'} + \bar{m}_\mu \iota^A o^{A'}.
\ee
It is easy to see that it is Hermitian. The metric (\ref{metric}) is then computed to be
\be
\eta_{\mu\nu} = - 2 l_{(\mu} n_{\nu)} + 2 m_{(\mu} \bar{m}_{\nu)}.
\ee
We also have the following expansion of the $\epsilon_{AB}$ symbol
\be
\epsilon_{AB}=o_A \iota_B -\iota_A o_B.
\ee

\subsection{Self-dual two-forms}

The following self-dual two-forms play the central role in the article. They are defined as
\be\label{Sigma-def}
\Sigma^{AB} = \frac{1}{2} \theta^{A}{}_{A'}\wedge \theta^{BA'}.
\ee
Explicitly, in terms of the null tetrad and the spinor basis we get
\be\label{Sigma}
\Sigma^{AB} = l\wedge m \, o^A o^B + \bar{m}\wedge n \, i^A i^B + (l\wedge n - m\wedge \bar{m}) i^{(A} o^{B)}.
\ee
 
\subsection{${\rm SU}(2)$ spinors}

We will need ${\rm SU}(2)$ spinors when we consider the Hamiltonian formulation of any of our fermionic theories. Our conventions here is reminiscent of those in Appendix A of  \cite{Ashtekar:1991hf}, but there are some differences. In particular, we use a Hermitian tetrad, while the convention in  \cite{Ashtekar:1991hf} is that the tetrad is anti-Hermitian. 

Let us first consider ordinary, non-Grassmann-valued spinors. To define ${\rm SU}(2)$ spinors we need a Hermitian positive-definite form on spinors. This is a rank 2 mixed spinor $G_{A'A}: \bar{G}_{A'A}=G_{A'A}$, such that for any spinor $\lambda^A$ we have $\bar{\lambda}^{A'} \lambda^A G_{A'A}>0$. Here $\bar{\lambda}^{A'}$ is the complex conjugate of $\lambda^A$, not to be confused with the Hermitian conjugate that is reserved for Grassmann-valued fields. We can define the ${\rm SU}(2)$ transformations to be those ${\rm SL}(2,\C)$ ones that preserve the form $G_{A'A}$. Then $G_{A'A}$ defines an anti-linear operation $\star$ on spinors via:
\be
(\lambda^\star)_A:=G^{AA'}\bar{\lambda}_{A'}.
\ee
We require that the anti-symmetric rank 2 spinor $\epsilon_{AB}$ is preserved by the $\star$-operation:
\be
(\epsilon^\star)_{AB}=\epsilon_{AB},
\ee
which implies the following normalisation condition
\be\label{G-norm}
G_{AA'} G^{A'}{}_{B}=\epsilon_{AB}.
\ee
Using the normalisation condition we find that $(\lambda^{\star\star})^A=-\lambda^A$ or
\be
\star^2=-1.
\ee
Thus, the $\star$-operation so defined is similar to a "complex structure", except for the fact that it is anti-linear:
\be
(\alpha \lambda^A +\beta \eta^A)^\star= \bar{\alpha} (\lambda^\star)^A+\bar{\beta} (\eta^\star)^A.
\ee
We note that using the $\star$-operation we can rewrite the positive-definite quantity $\bar{\lambda}^{A'} \lambda^A G_{A'A}$ as follows
\be
\bar{\lambda}^{A'} \lambda^A G_{A'A} = \lambda_A (\lambda^\star)^A  >0.
\ee

Now for the purpose of 3+1 decompositions to be carried out below, we need to introduce a special Hermitian form that arises once a time vector field is chosen. We can then consider the zeroth component of the soldering form
\be
\theta_0^{AA'}\equiv \theta_\mu^{AA'} \left( \frac{\partial}{\partial t}\right)^\mu = \frac{1}{\sqrt{2}} \left( o^A o^{A'} + \iota^A \iota^{A'}\right).
\ee
It is Hermitian, and so we can use a multiple of $\theta_0^{AA'}$ as $G^{AA'}$. It remains to satisfy the normalisation condition (\ref{G-norm}). This is achieved by
\be\label{herm-form}
G^{AA'} := \sqrt{2} \theta_0^{AA'}.
\ee
We then define the spatial soldering form via
\be\label{sigma-sp}
\sigma^{i\, AB}:= G^{AA'} \theta^{i\, B}{}_{A'},
\ee
which is automatically symmetric $\sigma^{i\,AB}=\sigma^{i\,(AB)}$ because its anti-symmetric part is proportional to the product of the time vector with a spatial vector, which is zero. Explicitly, in terms of the spinor basis introduced above we have
\be\label{sigma-small}
\sigma^{i\,AB}= -m^i o_A o_B +\bar{m}^i \iota_A \iota_B+\frac{z^i}{\sqrt{2}}(\iota_A o_B+o_A\iota_B ).
\ee
The action of the $\star$-operation on the basis spinors is as follows:
\be
(o^\star)^A =  \iota^A, \qquad (\iota^\star)^A =  -o^A.
\ee
It is then easy to see from (\ref{sigma-small}) that the spatial soldering form so defined is anti-Hermitian with respect to the $\star$ operation:
\be
(\sigma^{i\,\star})^{AB}=-\sigma^{i\,AB}.
\ee

The following property of the product of two spatial soldering forms holds:
\be\label{sigma-ident}
\sigma^i_A{}^B\sigma^j_B{}^C=\frac{1}{2}\delta^{ij} \epsilon_A{}^C-\frac{\im}{\sqrt{2}}\epsilon^{ijk} \sigma^k_A{}^C.
\ee
Below we will also often use the following related quantities 
\be\label{T}
T^i_A{}^B:=\im \sqrt{2} \sigma^i_A{}^B, 
\ee
which have the following nicer algebra:
\be\label{T-algebra}
T^i_A{}^B T^j_B{}^C=-\delta^{ij} \epsilon_A{}^C+ \epsilon^{ijk} \, T^k_A{}^C.
\ee

Now, using the Hermitian form (\ref{herm-form}), we extend the $\star$-operation defined above to Grassmann-valued spinors. Thus, we define a new operation on Grassmann-valued spinors which is a combination of the usual Hermitian conjugation $\dagger$ acting on a Grassmann-valued fermion with the operation of converting the primed index into an unprimed one:
\be
(\lambda^\star)^A := G^{AA'} (\lambda^\dagger)_{A'}.
\ee
This operation will be of importance when we discuss the 3+1 decomposition of the standard Weyl and Dirac actions.

\subsection{Hamiltonian description of a single massless Weyl fermion}

The 3+1 decomposition of (\ref{weyl-massless}) is given by
\be
{\cal L}_{\text{Weyl}}= \im \sqrt{2} (\lambda^\dagger)_{A'} \theta_0^{AA'} \partial_t \lambda_A -\im \sqrt{2} (\lambda^\dagger)_{A'} \theta^{i\,AA'} \partial_i \lambda_A.
\ee
It readily follows that the canonically conjugate momentum is given by
\be\label{pi-weyl}
\pi^A= \im \sqrt{2} (\lambda^\dagger)_{A'} \theta_0^{AA'} = \im (\lambda^\star)^A,
\ee
We note that the somewhat awkward factor of $\sqrt{2}$ in the original Lagrangian is needed precisely in order to have such a simple relation between the conjugate momentum $\pi^A$ and the $\star$-conjugate of $\lambda^A$. We can now rewrite our Lagrangian as
\be\label{weyl-ham-form}
{\cal L}_{\text{Weyl}}=\pi^A \partial_t \lambda_A - \im \pi^A T^i_A{}^B \partial_i \lambda_B,
\ee
where we have used the spatial soldering form in their version (\ref{T}). An alternative expression for the above Lagrangian is
\be
{\cal L}_{\text{Weyl}}=\im (\lambda^\star)^A \partial_t \lambda_A +(\lambda^\star)^A T^i_A{}^B \partial_i \lambda_B.
\ee
Using $(\lambda^A \eta^B)^\star = (\eta^\star)^{B} (\lambda^\star)^{A}$ as well as the fact that $\star^2=-1$ and that the quantities $T^i_A{}^B$ are $\star$-Hermitian, one can easily check this Lagrangian to be $\star$-Hermitian modulo a surface term.

A useful exercise for what follows is to find the field equations that follow from (\ref{weyl-ham-form}). Treating the fermionic fields $\lambda^A,\pi^A$ as independent we get:
\be
\dot{\lambda}_A-\im T^i_A{}^B\partial_i \lambda_B =0, \qquad
\dot{\pi}_A+\im T^i_A{}^B\partial_i \pi_B =0.
\ee
The second equation is the $\star$-conjugate of the first, as it should be. We can obtain a simpler second-order equation for $\lambda_A$ by differentiating its equation with respect to time, applying to it the operator $\im T^i_A{}^B\partial_i$, and then taking the difference of the results. We use the identity (\ref{T-algebra}) and get:
\be\label{box-lambda}
(\partial_t^2 -\partial_i\partial^i )\lambda_A=0,
\ee
which is the wave equation.

\subsection{Hamiltonian formulation of Majorana theory}

 It is easy to show that the last term in (\ref{majorana}) can be rewritten in terms of the momentum (\ref{pi-weyl}) as $-(m/2)\pi^A\pi_A$, and so the Lagrangian in the Hamiltonian form is:
\be\label{L-Maj-1}
{\cal L}_{\text{Majorana}} =\pi^A \partial_t \lambda_A - \im \pi^A T^i_A{}^B \partial_i \lambda_B -
(m/2)\lambda^A\lambda_A-(m/2)\pi^A\pi_A.
\ee
We can again rewrite this in terms of $\pi^A=\im(\lambda^\star)^A$ and then check that it is $\star$-Hermitian. Indeed, we have $(\pi^A \pi_A)^\star=\lambda^A\lambda_A$, so the last two terms go into each other under the $\star$-operation. Note that $\star^2=-1$ only when it acts on a fermionic quantity. Acting on a scalar this is simply the operation of complex conjugation.

Let us also give explicitly the Hamiltonian that corresponds to (\ref{L-Maj-1}). We have:
\be\label{H-Maj-orig}
{\cal H}_{\text{Majorana}} =(m/2)\pi^A\pi_A + \im \pi^A T^i_A{}^B \partial_i \lambda_B +
(m/2)\lambda^A\lambda_A.
\ee

Let us carry out the exercise of finding a second order differential equation for each field again. We have:
\be\label{majorana-eqs}
\dot{\lambda}_A-\im T^i_A{}^B\partial_i \lambda_B -m\pi_A=0, \qquad
\dot{\pi}_A+\im T^i_A{}^B\partial_i \pi_B +m\lambda_A =0.
\ee
As in the massless case, the two equations are the $\star$-conjugates of each other. One can find the momentum $\pi_A$ from the first equation and substitute the result to the second. Using (\ref{T-algebra}) and multiplying the result by $m$ one gets:
\be\label{box-m-eqn}
(\partial_t \partial_t - \partial_i\partial^i +m^2 )\lambda_A =0,
\ee
which is the desired massive wave equation for a two-component fermion.

\end{document}